\documentclass[aps,prd,preprint,superscriptaddress,nofootinbib,showpacs]{revtex4}
\usepackage{graphics}
\usepackage{epsfig}
\usepackage{latexsym}
\usepackage{colordvi}
\usepackage{amsmath}
\usepackage{amssymb}
\newcommand{\be}{\begin{equation}}
\newcommand{\ee}{\end{equation}}

\begin{document}
\preprint{\parbox[b]{1in}{ \hbox{\tt PNUTP-07-A05, PNU-NTG-08/2007 } \hbox{\tt KIAS-P07039 } }}

\title{The Electric Dipole Moment of the Nucleons in Holographic QCD}

\author{Deog Ki Hong}
\email[E-mail: ]{dkhong@pusan.ac.kr} \affiliation{Department of
Physics, Pusan National University,
             Busan 609-735, Korea}
            % \affiliation{Asia Pacific Center for Theoretical Physics,  POSTECH, Pohang 709-784, Korea}

\author{Hyun-Chul Kim}
\email[E-mail: ]{hchkim@pusan.ac.kr} \affiliation{Department of
Physics, Pusan National University,
             Busan 609-735, Korea}
\affiliation{Nuclear Physics \& Radiation Technology Institute
(NuRI), Pusan National University,
             Busan 609-735, Korea}

\author{Sanjay Siwach}
\email[E-mail: ]{sksiwach@hotmail.com} \affiliation{Department of
Physics, Pusan National University,
             Busan 609-735, Korea}

\author{Ho-Ung Yee}
\email[E-mail: ]{ho-ung.yee@kias.re.kr} %\\
\affiliation{School of Physics, Korea Institute for Advanced
Study, Seoul 130-012, Korea} \vspace{0.1in}

\date{\today}

\begin{abstract}
We introduce  the strong CP-violation in the framework of AdS/QCD
model and calculate the electric dipole moments of nucleons as
well as the CP-violating pion-nucleon coupling. Our holographic estimate
of the electric dipole moments gives
for the neutron
$d_n=1.08\times 10^{-16}\,\, \bar\theta\, e\cdot {\rm cm}\,$,
which is
comparable with previous estimates.
We also predict that the electric dipole moment
of the proton should be precisely the minus of the neutron electric dipole moment,
thus leading to a new sum rule on the electric dipole moments of baryons.
\end{abstract}
%%% 11.25.Tq Gauge/string duality
%%% 11.10.Kk Field theory in dimension other than four
%%% 14.20.Dh properties of neutrons and protrons
%%% 11.25.Wx String and brane phenomenology
%%% 12.38.Cy Summation of perturbation theory in QCD
%%% 12.60.Fr Extension of electroweak Higgs sector
%%% 12.15.Ji Applications of electroweak models to specific processes
%%% 12.15.Lk Electroweak radiative corrections
%%% 12.60.Nz Technicolor models

\maketitle

\newpage

\section{Introduction}
The strong interactions of elementary particles are known to be
highly symmetric. The most stringently tested global symmetry of strong interactions
in the framework of relativistic quantum field theory is CP, charge
conjugation (C) times parity (P), or the time reversal symmetry, T.
The experimental upper bound on the CP violation comes from the
absence of the electric dipole moment (EDM) of neutron,
\begin{equation}
|d_n|< 2.9\times 10^{-26}\, e \cdot\mathrm{cm}\;\;
\end{equation}
with $90\,\%$ confidence level~\cite{Baker:2006ts}. More precise
measurements of the neutron EDM are now under way to improve the
current limit by a factor of 50 to 100~\cite{EDM}. On the other
hand, the standard theory of the strong interactions, QCD, allows a
CP violating term, called the $\theta$ term,
\begin{equation}
{\cal L}_{\rm
QCD}\ni\frac{\theta}{64\pi^2}\,\epsilon_{\mu\nu\rho\sigma}G^{a\mu\nu}G^{a\rho\sigma}\,,
\end{equation}
where $G_{\mu\nu}^a$ is the field strength tensor of gluons.

The quark mass term in the QCD Lagrangian,
\begin{equation}
{\cal L}_m = -\bar q_L^i(M_q)_{ij} q_R^j +{\rm h.c.}\quad,
\end{equation}
shifts $\theta$ by a chiral rotation,
%\begin{equation}
$\bar{\theta} = \theta + \mathrm{Arg\,Det} M_q$,
%\end{equation}
which is the physical strong CP-violation angle. The $\bar\theta$
term contributes to the neutron electric dipole
moment~\cite{Baluni:1978rf,Crewther:1979pi,Kawarabayashi:1980uh,Kim:1986ax,Shintani:2006xr,Pich:1991fq,Pospelov:1999ha}
\begin{equation}
d_n=c\times 10^{-16}\,\bar\theta\,e \cdot\mathrm{cm}\,,
\end{equation}
where $c$ is a constant of order one, as can be seen from the naive dimensional analysis,
\be d_n\sim \left({1\over m_N}\right)\left({m_q\, \bar\theta\over m_N}\right)\,\ee
with $m_N$ and $m_q$ being the masses of nucleons and  quarks, respectively.
The $\theta$ parameter
therefore has to be extremely fine-tuned to be consistent with the
experimental data,
\begin{equation}
\theta + \mathrm{Arg\,det} M\lesssim10^{-9}\,.
\end{equation}
Such fine-tuning is known as the strong CP problem and several
solutions are proposed to solve the strong CP
problem~\cite{Peccei:2006as}.

Since not only QCD but the electroweak
interactions~\cite{Georgi:1986kr} and also the physics beyond the
standard model contribute to the neutron electric dipole moment, it
is quite important to estimate the QCD contribution accurately. In
this letter we estimate the electric dipole moment of nucleons as
well as the CP-violating pion-nucleon coupling in holographic models
of QCD, which have been quite successful in describing the
properties of hadrons. We find that our holographic estimate of
nucleon EDM is comparable with previous results, based on lattice
calculations~\cite{Shintani:2006xr}, current
algebra~~\cite{Baluni:1978rf,Crewther:1979pi,Kawarabayashi:1980uh},
chiral perturbation theory~\cite{Pich:1991fq} or QCD sum
rule~\cite{Pospelov:1999ha}. We also get an interesting sum rule for
the nucleon EDM's; the EDM of neutron is opposite to that of proton,
$d_n+d_p=0$, which is consistent with the recent lattice
result~\cite{Shintani:2006xr} and the estimate from the Light-Front
formalism~\cite{Brodsky:2006ez}.

The new sum rule on EDM is insensitive to any higher order corrections in
$1/N_c$ and is also a model-independent prediction of holographic
QCD, where baryons are realized as instanton solitons.
As was shown in~\cite{HRYY}, the Pauli term in the 5D action
should not have any $U(1)$ coupling, since the instanton has only $SU(2)$ nonabelian
tails, and thus the anomalous magnetic moments of baryons should add up to zero
for each flavor multiplets. The same should hold for the electric dipole moments,
because they are related to the anomalous magnetic moments by a $U(1)$ axial rotation.

\section{The Model with Baryons}

A holographic model for spin $1\over 2$ baryons is constructed for
two flavors in~\cite{Ho}.\footnote{See Ref.\cite{de
Teramond:2005su} for a model of higher spin Regge trajectory.} In
this section, we briefly summarize the model~\cite{Ho,Kim:2007xi},
since we will be studying the nucleon EDM in the framework of this
model, closely following the notations.
%Throughout this paper, we will focus on the $N_F=2$ sub-sector of the full QCD, as it is
%justified for low-energy chiral dynamics.
For the meson sector, we
take the simplest hard-wall AdS/QCD model as in~\cite{EKSS,PR}, with the metric
\be
ds^2 = {1\over z^2} \left(-dz^2 + \eta_{\mu\nu}dx^{\mu}dx^{\nu}\right)\quad,
\ee
where $0 \le z \le z_m$ and $\eta_{\mu\nu}={\rm diag}(+1,-1,-1,-1)$.
This model captures the important aspects of low energy
chiral dynamics of light mesons, especially that of pions and vector mesons.
One should think of this type of holographic models as alternative effective
theories of strongly coupled field theory like low-energy QCD in the large $N_c$ limit.
The theory is expected to have a classical nature in the large $N_c$ approximation.
In conjunction with the renormalization group invariance, which
is an essential element of quantum field theories, the resulting large $N_c$ classical master field
should develop a new, dynamically generated, space, which corresponds to the energy scale or
the renormalization scale of the Wilsonian type.
The large $N_c$ classical nature and
the Wilsonian renormalization group fit together in the extra dimension.
%This seems to be the  way of compromising between the large
The low-lying spin $1\over 2$ baryons such as
a proton-neutron isospin doublet and its excitations are shown to be naturally realized as
5D Dirac spinors in this picture~\cite{Ho}.

Since we are interested in the spin $1\over 2$, isospin $1\over 2$ baryons,
we introduce (Dirac) spinors in our 5 dimensional AdS slice as a holographic
realization of spin $1\over 2$ baryons.\footnote{Note that
a 5D spinor has 4-components, like a Dirac spinor in 4D.}
However, there are two caveats we have to be careful about.
The first one is the representation of our 5D holographic baryon fields under the
chiral symmetry $SU(2)_L\times SU(2)_R$ of QCD, which becomes a gauge symmetry in the dual 5D model.
We know that the
lowest-lying 4D excitations, the nucleons, form a doublet under the isospin $SU(2)_I$,
the diagonal part of $SU(2)_L\times SU(2)_R$, after chiral symmetry breaking,
but there is no unique way to assign the nucleon charges under
the original $SU(2)_L\times SU(2)_R$ symmetry, since the Nambu-Goldstone fields can be
always multiplied to the nucleon fields~\cite{Georgi:1985kw}.

For $N_F=2$ case however, there is an answer. To match the UV
anomaly of $SU(2)_L\times SU(2)_R\times U(1)_B$ from massless
chiral quarks $(u_L,d_L)$ and $(u_R,d_R)$ with $N_c=3$, there must
exist massless chiral baryon doublets $(p_L,n_L)$ and $(p_R,n_R)$
with the representations $(\square,1)$ and $(1,\square)$ under
$SU(2)_L\times SU(2)_R$ respectively, when the theory is confining
but in the chirally symmetric false vacuum~\cite{'t Hooft:1979bh}.
Their baryon charge is $N_c=3$ times the quark baryon charge.

Chiral symmetry breaking in the true vacuum introduces a mass
coupling for nucleons, \be {\cal L}_{\chi SB}\sim
-m_N\left(\begin{array}{c}\bar p_L\\\bar n_L\end{array}\right)
\Sigma \left(\begin{array}{cc} p_R & n_R\end{array}\right) + {\rm
h.c.}\quad,\label{massinbroken} \ee where $\Sigma=\exp({2i
\pi\over f_\pi})\in SU(2)$ is the non-linear group field in the
broken phase, which transforms non-linearly as $\Sigma \to U_L\,
\Sigma\, U_R^\dagger$ under $SU(2)_L\times SU(2)_R$. As $\left<
\Sigma \right>={\bf 1}$ in the true vacuum, the above is invariant
under the isospin $SU(2)_I$ for which nucleons form a doublet. In
the symmetric (false) vacuum, $\Sigma$ and the isospin singlet the
sigma meson ($\sigma$) will be completed to a linear field $X$
which is bi-fundamental $(\square,\bar\square)$ under
$SU(2)_L\times SU(2)_R$, and we should have the following term in
the theory \be {\cal L}_m = - g \left(\begin{array}{c}\bar
p_L\\\bar n_L\end{array}\right) X \left(\begin{array}{cc} p_R &
n_R\end{array}\right) + {\rm h.c.}\quad,\label{nucleonmass} \ee to
have the coupling (\ref{massinbroken}) in the broken phase $\left<
X\right> \sim \Lambda_{\rm QCD} {\bf 1}$.

Based on the above consideration, we find the simplest choice is
to introduce two 5D spinors $N_1$ and $N_2$, of representation
$(\square,1)$ and $(1,\square)$ respectively under $SU(2)_L\times SU(2)_R$ 5D gauge symmetry~\cite{Ho}.
Upon the Kaluza-Klein (KK)
reduction to 4D, the modes from $N_1$ and $N_2$ must include the above-mentioned
massless chiral baryon excitations $(p_L,n_L)$ and $(p_R,n_R)$ respectively, in the unbroken chiral-symmetric limit.
This requirement uniquely fixes the IR boundary conditions for $N_1$ and $N_2$ at $z=z_m$.
Then, the natural holographic realization of the nucleon mass coupling (\ref{nucleonmass})
is to introduce a gauge invariant 5D interaction
\be
{\cal L}_{\rm 5D}\,\, =\,\, -g \,\bar N_1 X N_2 \,\,+\,\, {\rm h.c.}\,,
\ee
where $X$ is the bi-fundamental 5D scalar field of $(\square,\bar\square)$, whose
vacuum expectation value (VEV)
breaks the chiral symmetry in the model. The above coupling
will induce the mass coupling (\ref{massinbroken}) between the would-be massless 4D chiral baryons
from $N_1$ and $N_2$ in the chiral-symmetry broken vacuum $X\sim {1\over 2}\Sigma z^3$.
The coupling strength $g$ must then be fitted to reproduce the nucleon mass $m_N=0.94$ GeV
as the lowest mass eigenvalue.

The other caveat in our holographic baryon model is
the question of chirality in the 5 dimensional context.
Though $N_1$  ($N_2$) is the holographic dual field  to the 4D left(right)-handed
nucleon operator,\footnote{
We are using the same notation for both the operator and the state of the 4D chiral baryons without confusion.
Strictly speaking, $N_1$ is dual to the left-handed chiral operator made of three left-handed quarks, and
vice versa for $N_2$.}
there is no chirality in 5 dimensions.
The 4D chirality is in fact encoded in the sign of 5D Dirac mass
term \cite{Henningson:1998cd}. For a positive 5D mass,\footnote{Our convention is ${\cal L}_m=-m_5 \bar N N$ and
$\Gamma^5=(-i)z\gamma^5$ with $\gamma^5 \psi_L=+\psi_L$.}
 only the right-handed component of the 5D spinor
survives near the boundary $z\to 0$, and this acts as a source for the left-handed chiral operator in 4 dimensions.
The story is simply reversed for the opposite sign case.
The magnitude of the 5D mass is given by the AdS/CFT relation
\be
m_5^2 \,\,=\,\,(\Delta-2)^2\,,
\ee
where we  take $\Delta={9\over 2}$ for a composite baryon operator of three quarks.
Considering the anomalous dimension of the composite operator might lead to slightly
different results. %would be an important future direction.

To wrap up the above discussions,
our 5 dimensional holographic model for spin $1\over 2$, isospin $1\over 2$
baryons in $N_F=2$ sector is given by
\begin{eqnarray}
S_{\rm kin}& =& \int dz \int dx^4 \sqrt{G_5}\,\left[i \bar N_1 \Gamma^M D_M N_1+i \bar N_2 \Gamma^M D_M N_2
-{5\over 2}\bar N_1 N_1
+{5\over 2}\bar N_2 N_2\right]\quad,\nonumber\\
S_m &=& \int dz \int dx^4 \sqrt{G_5}\,\left[ -g \bar N_1 X N_2
-g \bar N_2 X^\dagger N_1\right]\,,\label{action}
\end{eqnarray}
where $D_M$  is the gauge and Lorentz covariant derivative,
$\sqrt{G_5}={1\over z^5}$, and  the gamma matrices
in our ${ AdS}_5$ are related to the 4D gamma matrices as
$\Gamma^{\mu}=z \gamma^{\mu}$ for $\mu=0,1,2,3$ and $\Gamma^5=-iz\gamma^5$.
Upon KK reduction to 4D, it is easy to find the
eigenmode equations for the mass spectrum of 4D spin $1\over 2$ baryons.
Writing $N_1(x,z)=f_{1L}(z)B_L(x)+f_{1R}(z)B_R(x)$ and similarly
for $N_2(x,z)=f_{2L}(z)B_L(x)+f_{2R}(z)B_R(x)$, where $B_{L,R}$ are
the components of the 4D eigenmode spinor $B=(B_L, B_R)^T$ with mass $m_N$ to be determined, we have
\begin{eqnarray}
&&\left(
\begin{array}{cc}
\partial_z -{\Delta \over z} & -{g \left<X\right>\over z} \\
-{g \left<X^\dagger\right>\over z} & \partial_z -{4-\Delta \over z}
\end{array}
\right)
\left(
\begin{array}{c} f_{1L} \\ f_{2L}
\end{array}
\right)=-m_N
\left(
\begin{array}{c} f_{1R} \\ f_{2R}
\end{array}
\right)\,, \nonumber\\
&&\left(
\begin{array}{cc}
\partial_z -{4-\Delta \over z} &{g \left<X\right>\over z}\\
{g \left<X^\dagger\right>\over z} & \partial_z -{\Delta \over z}
\end{array}
\right)
\left(
\begin{array}{c} f_{1R} \\ f_{2R}
\end{array}
\right)=m_N
\left(
\begin{array}{c} f_{1L} \\ f_{2L}
\end{array}
\right)\,,
\label{kk_mode}
\end{eqnarray}
with $\Delta$ ($={9\over 2}$ in our case) in general.
As mentioned before, the existence of the chiral zero modes when $\left<X\right>=0$ requires us
the IR boundary condition  $f_{1R}(z_m)=f_{2L}(z_m)=0$.
In the meson sector the best fit was found
for $\left<X\right>={1\over 2}(m_q z+\sigma z^3)$
with $m_q=2.34$ MeV, $\sigma=(311 \,{\rm MeV})^3$,
and the IR cut-off $z_m=(330\, {\rm MeV})^{-1}$~\cite{EKSS,PR}.
Then, the only remaining parameter of the theory is the dimensionless coupling $g$,
which was found to be $g=9.18$ in~\cite{Ho} to reproduce $m_N=0.94$ GeV as a lowest mass eigenvalue.

In~\cite{HRYY}, which discussed nucleons in the top-down
Sakai-Sugimoto model, an important new operator,
responsible for anomalous magnetic dipole moments, among others, was identified,
which will also be important for our analysis of CP-violating
electric dipole moments. Experimentally, the proton magnetic
moment is $\mu_p=2.8\mu_N=\mu_N+1.8\mu_N$ where the Nuclear
Magneton $\mu_N={e\over 2 m_N}$ is from the minimal coupling of
charge 1, while the neutron has $\mu_n=-1.8\mu_N$. It is clear
that the anomalous piece has a structure of isospin doublet
without $U(1)_B$ charge, as shown in~\cite{HRYY}. Since electromagnetic coupling is a sum
of the isospin and $U(1)_B$, $Q={1\over 2}B+I_3$,
the operator responsible for the anomalous magnetic moment does
not include $U(1)_B$.

It is easy to find the corresponding 5D operator which induces the desired anomalous magnetic moment
upon 4D reduction,
\be
{\cal L}_{\rm dipole}=i D \left[ \bar N_1 \Gamma^{MN}(F_L)_{MN}N_1 - \bar N_2 \Gamma^{MN}(F_R)_{MN}N_2\right]\quad,
\label{5Ddipole}
\ee
where $F_{L,R}$ are the field strengths of $A_{L,R}$ and $D$ is a real parameter that
must be fixed to reproduce $\mu_{\rm anomalous}=1.8 \mu_N$.\footnote{In~\cite{HRYY}, $D$
was reliably predicted from string theory, which explains $\mu_{\rm anomalous}=1.8 \mu_N$ quite well.
But in our bottom-up model it is a fitting parameter, though
the relative minus sign in (\ref{5Ddipole})
is dictated by 4D parity invariance, as easily seen from the fact that
5D Dirac mass term flips its sign under 4D parity.}
Note that $A_{L,R}$ in our model do not include $U(1)_B$, as required.
In fact, this absence of $U(1)_B$ in the 5D effective operator (\ref{5Ddipole})
has its origin from the fact that baryons in AdS/QCD arise as instantonic solitons
of small size in our (4+1)-dim gauge theory of $A_{L,R}$.
As we treat them as point-like with the effective fields $N_1$ and $N_2$,
we need to take into account their long-range instanton tail of $A_{L,R}$, and
the operator (\ref{5Ddipole}) exactly sources those tail profiles.
In a more complete description, the coefficient $D$ would be determined
by the stabilized size of small instanton-solitons, whereas here it is a fitted parameter against
experiments. Since instanton profiles are purely non-abelian, the operator (\ref{5Ddipole})
should not include $U(1)_B$, which leads to a model-independent sum rule
for any quantities derived from the operator such as the anomalous magnetic moments.

\section{Physics of Vacuum Alignment}

 The strong CP-violation in QCD can be introduced either in the vacuum $\theta$-angle
or as complex phases in the quark mass matrix $M_q$ defined as
\be
{\cal L}_m = -\bar q_L^i(M_q)_{ij} q_R^j +{\rm h.c.}\quad.\label{quarkmass}
\ee
As is well-known, the anomalous axial $U(1)_A$ rotation can shift the $\theta$-angle
to zero, making the strong CP-violation appear only in $M_q$ or vice versa.
The physical strong CP-violation angle is\footnote{The $\theta$-angle
is normalized as ${\cal L}_\theta = {\theta \over 32 \pi^2}{\rm Tr} (F\wedge F)$ with
$ {1 \over 32 \pi^2}\int {\rm Tr}(F\wedge F )=1$ for a single QCD instanton.}
 \be
 \bar\theta = \theta + {\rm Arg}\,{\rm Det}(M_q)\quad.
 \ee
Presumably, the $\theta$-angle can be described in the holographic model
by a kind of axion~\cite{Katz:2007tf},
  but for our purpose it is much more convenient
to work in the frame where the strong CP-violation appears in
$M_q$ only. This is because in our AdS/QCD model, $M_q$ is easily
described by the non-normalizable mode of the scalar $X$, \be
\left<X(z)\right>={1\over 2}M_q z +{1\over 2}\Sigma z^3 \,, \ee
where $\Sigma$ is the quark bi-linear condensate
$(\Sigma)^{ij}=-\left<\bar q_R^j q_L^i\right>$.

In the CP-symmetric case of $M_q={\rm diag}(m_u, m_d)$ with small real masses $m_u,m_d\ll \Lambda_{\rm QCD}$,
it looks natural to have the bi-quark condensate proportional to identity matrix\footnote{Isospin
violating effects from $m_u\ne m_d$ to the bi-quark condensate can be neglected to first order
in $m_q$, as we are interested in the results that are first order in $m_q$ and $\bar \theta$.}
\be
\Sigma \sim \Lambda^3_{\rm QCD} {\bf 1}_2\quad,
\ee
as was used before and will be shown rigorously in a moment.
However, as we turn on small CP-violation as phases $\alpha_{1,2}$ of quark mass
$M_q={\rm diag}(e^{i\alpha_1} m_u, e^{i\alpha_2} m_d)$,
it is not obvious which polarization $\Sigma$ will take in the $SU(2)$ space
of its moduli, $\Sigma=\Lambda^3 U$ with $U\in SU(2)$ and
$\Lambda$ is a scale proportional to $\Lambda_{\rm QCD}$, the intrinsic scale of QCD.
This is the problem
of vacuum alignment, first considered by Dashen~\cite{Dashen:1970et}.

The vacuum moduli of $\Sigma$
form a coset space  $SU(2)_L\times SU(2)_R/SU(2)_I$.
Since the quark mass $M_q$ breaks the chiral symmetry, it lifts
the moduli space by inducing a potential,
\be
V_m =-{\rm Tr}(M_q \Sigma^\dagger)+{\rm h.c.}\quad,
\ee
which can be inferred from (\ref{quarkmass}).
With our diagonal mass matrix $M_q$,
it is natural to make $\left<\Sigma\right>$ align along its diagonal direction
by a $SU(2)_L\times SU(2)_R$ chiral transformation, that is,
\be
\left<\Sigma\right>=
\Lambda^3\left(\begin{array}{cc} e^{ix} & 0\\ 0& e^{-ix}\end{array}\right)\quad,
\ee
to get the vacuum potential $V_m=-2m_u\,\Lambda^3\,\cos(x-\alpha_1) -2m_d\,\Lambda^3\,
 \cos(x+\alpha_2)$.
For small $\alpha_{1,2}$, related to $\bar\theta\sim 10^{-9}$, as will be
seen in a moment,
we expand the above cosines to get $V_m/\Lambda^3
\approx (m_u+m_d)x^2 -2(m_u\alpha_1-m_d\alpha_2)x+{\rm const.}$
The potential minimizes at
\be
x={m_u\alpha_1-m_d\alpha_2 \over m_u+m_d}\quad,
\ee
as the aligned vacuum polarization in the presence of the strong CP-violation. It also
confirms that identity matrix is the ground state for the CP-symmetric case,
$\alpha_{1,2}=0$\,.

For our analysis, it is more convenient to take a non-anomalous $SU(2)_A$ axial rotation
to remove the phase angle $x$ in $\Sigma$.\footnote{Note that this global rotation also
rotates $M_q$ simultaneously. When we discussed the lifted moduli of $SU(2)$ above, we fixed $M_q$.
In other words, only the relative $SU(2)_A$ angle between $M_q$ and $\Sigma$ is physical and
we can always go to the frame where $\Sigma\sim {\bf 1}$, as the coset space is homogeneous.}
This brings the quark mass angles to $\alpha_1\to \alpha_1-x={(\alpha_1+\alpha_2) m_d\over m_u+m_d}$ and
$\alpha_2\to \alpha_2+x={(\alpha_1+\alpha_2)m_u \over m_u+m_d}$.
Since the physical CP-violation angle is $\bar\theta=(\alpha_1+\alpha_2)$, we
finally end up with the quark mass matrix
\be
M_q = \left(\begin{array}{cc} m_u \,e^{i\bar\theta( { m_d \over m_u+m_d})} & 0 \\
                   0 &  m_d\, e^{i \bar\theta({m_u \over m_u+m_d})}\end{array}\right)
                   \quad {\rm and}\quad \left<\Sigma\right>=\sigma {\bf 1}\,.
\label{cpoddmass}
\ee
%and $\left<\Sigma\right>=\sigma {\bf 1}\sim \Lambda_{\rm QCD}^3 {\bf 1}$ as the background.
Therefore, in our AdS/QCD model the strong CP-violation is easily
encoded in the VEV of $X$ as $\left< X(z)\right>={1\over 2}\left(M_q z+\left<\Sigma\right> z^3\right)$
with
$M_q$ and $\left<\Sigma\right>$ given in~(\ref{cpoddmass}). This is the starting point
for our holographic analysis of the strong CP-violation.

Note that we introduce
strong CP-violation only through the VEV of $X$, while keeping the AdS/QCD theory itself,
namely the various couplings and all other parameters of AdS/QCD, CP-conserving, because
  the QCD dynamics  is CP-conserving when the angle $\bar\theta$ vanishes, and  %, which
the angle can be absorbed into ``external" quark mass parameters.

As we are interested in the effects of the strong CP violation in the first order of
$\bar\theta$ (and $m_{u,d}$),
we expand the phase exponents in $M_q$ to get
\be
M_q \simeq\left(\begin{array}{cc} m_u  & 0 \\
                   0 &  m_d\end{array}\right)+i \,\bar\theta\, \bar m\,{\bf 1}=M_q^0+i\,\delta M_q\quad,
\ee
where $\bar m^{-1}=m_u^{-1}+m_d^{-1}$ is the reduced mass.
The CP-violating axial mass $i\,\delta M_q$
is proportional to the identity matrix~\footnote{The angle $\bar\theta$ should be proportional to
$\gamma_5$, which is suppressed here, since we are using the chiral basis.},
which could also be argued from the vacuum
stability~\cite{Baluni:1978rf,Crewther:1979pi}.
This in fact agrees with the previously known chiral Lagrangian with the strong CP-violation.
For simplicity, we will take an isospin symmetric quark mass $m_u=m_d=m=2\bar m$, then
the whole $\left<X\right>$ is proportional to the identity matrix, which we write as
\be
\left<X(z)\right>=\left[{1\over 2}(mz+\sigma z^3)+{i\over 4}m\bar\theta \,z \right]{\bf 1}
=v(z){\bf 1}=\left[v_0(z)+i\delta v(z)\right]{\bf 1}\,,
\ee
where $v_0(z)$ is the dominant CP-conserving piece
while small $i\delta v(z)=i{m\bar\theta \over 4}z$ breaks the CP symmetry.

\section{The Neutron Electric Dipole Moment}

The effects of the strong CP-violation in the VEV of $X$
are generically of two kinds in the holographic models; it modifies  the wave functions of
4D excitations along the fifth direction if the corresponding 5D field couples to $\left<X\right>$, or
it affects various couplings of 4D interactions after integrating over the fifth direction
if the 5D coupling involves $X$. In consideration of the latter, we should also
take into account the former as well.
The CP-violating pion-nucleon coupling $\bar g_{\pi NN}$
that we will calculate in section \ref{secv} is such an example.

 The operator
(\ref{5Ddipole}) will be responsible for the electric dipole moment of the nucleons,
as well as the magnetic dipole moment.
This is because an axial $U(1)_A$ rotation of the magnetic dipole moment
is precisely the electric dipole moment, and the strong CP-violation can be
described by the non-zero $U(1)_A$ phases of the quark mass matrix, so that
its effect will appear as a small $U(1)_A$ rotation of the magnetic dipole moment.
We stress that the 5D operator (\ref{5Ddipole}) itself is CP-conserving, and we don't
include explicit CP-violating 5D operator in the model. After introducing the CP-violation
in $\left<X\right>$ in the otherwise CP-conserving theory, additional CP-violating
5D operators may be generated by 5D loop effects. However, in the spirit of AdS/QCD
we keep the tree-level analysis only, systematically ignoring $1\over N_c$ corrections.
Therefore, we expect our analysis captures a leading $N_c$ contribution only except
the sum rule, $d_p+d_n=0$, whose origin goes beyond the $1/N_c$ approximation.
 The CP-violation introduced
by $\left<X\right>$  affects the 5D wave functions $f_{(1,2)(L,R)}$
of the (lowest) nucleon state through the mass coupling
\be
{\cal L}_m = -g\bar N_1 \left<X\right> N_2 + {\rm h.c.}\quad,
\ee
and the resulting 4D operator of nucleons after integrating over the fifth coordinate with these
wave functions
will have a small CP-violating counterpart of the magnetic dipole moment, which is the electric dipole moment
we are interested in.

To be explicit, the CP-violating $\left<X\right>=v(z){\bf 1}$ enters  the
 equation for the nucleon wave function $N_1=f_{1L}(z)B_L(x)+f_{1R}(z)B_R(x)$
 and  $N_2=f_{2L}(z)B_L(x)+f_{2R}(z)B_R(x)$, where $B_{L,R}(x)$ is the 4D nucleon field, as follows
\begin{eqnarray}
\left(\begin{array}{cc} \partial_z-{\Delta\over z} & -g{ v(z)\over z}\\
                       -g {v(z)^\dagger\over z} & \partial_z -{4-\Delta \over z}\end{array}
                       \right)\left(\begin{array}{c}f_{1L}\\f_{2L}\end{array}\right) & = &
                       -m_N \left(\begin{array}{c} f_{1R}\\ f_{2R}\end{array}\right)\nonumber\\
\left(\begin{array}{cc} \partial_z-{4-\Delta\over z} & g{v(z)\over z}\\
                       g{v(z)\dagger\over z} & \partial_z -{\Delta \over z}\end{array}
                       \right)\left(\begin{array}{c}f_{1R}\\f_{2R}\end{array}\right) & = &
                       m_N \left(\begin{array}{c} f_{1L}\\ f_{2L}\end{array}\right)\quad,\label{eqn}
\end{eqnarray}
where $\Delta={9\over 2}$ is the scaling dimension of the 4D baryon operator and $m_N=0.94$ GeV
is the nucleon mass.
To get the standard 4D kinetic term for $B(x)$, we have to normalize the eigen-functions
\be
\int_0^{z_m }dz\left[{1\over z^4}\left(|f_{1L}|^2+|f_{2L}|^2\right)\right]=
\int_0^{z_m} dz\left[{1\over z^4}\left(|f_{1R}|^2+|f_{2R}|^2\right)\right]=1\quad.
\ee
We are interested in the effects of small CP-violating imaginary part in $v(z)=v_0(z)+i\delta v(z)$
to linear order.
The CP-conserving zeroth order eigenfunctions $f^{(0)}_{(1,2)(L,R)}$ with $v_0(z)$,
obtained numerically in~\cite{Ho},
are real functions with an important property
\be
\left(\begin{array}{c} f_{1L}^{(0)} \\ f_{1R}^{(0)}\end{array}\right)=\left(\begin{array}{cc}
0 & 1\\ -1 & 0\end{array}\right)\left(\begin{array}{c} f_{2L}^{(0)}\\f_{2R}^{(0)}\end{array}\right)\quad,
\label{wave}
\ee
which can also be  checked explicitly from (\ref{eqn}).
This is a simple consequence of 4D parity. The precise parity transformation
in our model that leaves the 5D action invariant is
\begin{eqnarray}
(x_0,\vec x,z) & \longleftrightarrow & (x_0,-\vec x,z)\nonumber\\
(A_L^0,\vec A_L,A_L^z) & \longleftrightarrow & (A_R^0,-\vec A_R,A_R^z)\nonumber\\
\left(\begin{array}{c}
N_1\\
N_2
\end{array} \right)& \longleftrightarrow  &\left(\begin{array}{cc}
0 & \gamma^0\gamma^5\\ -\gamma^0\gamma^5 & 0\end{array}\right)\left(\begin{array}{c}
N_1\\
N_2
\end{array} \right)\,.
\end{eqnarray}
%where the last one is in the chiral basis $N_{i}=(N_{iL}, N_{iR})^T$.
We see that the (lowest) nucleon state
(\ref{wave}) is parity even as expected. Using this and the
explicit form of $\delta v(z)={m\,{\bar\theta}}\,z/4$, it is not
difficult to find that the first order effect of $i\delta v(z)$ on
the wave functions is a simple phase factor in the eigenfunctions
$f_{(1,2)(L,R)}$, \be \left(\begin{array}{c} f_{1L}\\
f_{2L}\end{array}\right)=e^{i\alpha} \left(\begin{array}{c}
f_{1L}^{(0)}\\ f_{2L}^{(0)}\end{array}\right)\quad, \quad
\left(\begin{array}{c} f_{1R}\\
f_{2R}\end{array}\right)=e^{i\beta} \left(\begin{array}{c}
f_{1R}^{(0)}\\ f_{2R}^{(0)}\end{array}\right)\quad, \ee with \be
(\alpha-\beta)={g\over m_N}\left(\delta v(z)\over z\right)={g
\over 4}{m\over m_N}\,\bar\theta\quad, \label{angle} \ee without
affecting the mass $m_N=0.94$ GeV. Note that the common phase on
$f_L$ and $f_R$ is not physical as it wouldn't appear in any
bilinear operators $\bar N \Gamma^{MN\cdots} N$, while the above
difference in phases (\ref{angle}) is a physical CP-violating
effect. It is also possible to work in the frame where all wave
functions are real, and the mass $m_N$ instead gets a phase
$m_N\to e^{i(\alpha-\beta)} m_N$,\footnote{ In the general complex
mass case, the mass $m_N$ in the first equation in (\ref{eqn})
should be replaced by $m_N^*$.} although we will stick to the
previous description for a while. We will get back to the frame of
complex mass and real wave functions later, when we argue the
generality of our result against other possible higher dimensional
operators.

Inserting the expansion $N_1=f_{1L}(z)B_L(x)+f_{1R}(z)B_R(x)$
 and  $N_2=f_{2L}(z)B_L(x)+f_{2R}(z)B_R(x)$ into our relevant 5D operator (\ref{5Ddipole}),
 \be
 S_{\rm dipole} = iD \int d^4x\int dz\,\sqrt{G_5}\,
 \left[ \bar N_1 \Gamma^{MN}(F_L)_{MN}N_1 - \bar N_2 \Gamma^{MN}(F_R)_{MN}N_2\right]\,,
 \ee
 it is straightforward to obtain the resulting 4D dipole moment operators.
 To read off the electromagnetic coupling, recall that $Q=I_3+\cdots$ with
 the isospin $I_3=(t_3^L+t_3^R)$, so that we can simply replace $A_{L,R}$ by
 $A_L=e A^{\rm em}t_3$ and $A_R=e A^{\rm em}t_3$ with $t_3=\frac12\sigma_3$
 to extract couplings to the electromagnetic
 vector potential $A^{\rm em}$. According to the AdS/CFT correspondence,
 the external vector potential is simply a non-normalizable mode of the
 corresponding 5D gauge field, which turns out to be a simple constant mode over the fifth direction.

A quick calculation results in the following 4D dipole moment operators,\footnote{Our convention is
$\sigma^{\mu\nu}=i\gamma^{\mu\nu}={i\over 2}[\gamma^\mu,\gamma^\nu]$ with
$\gamma^\mu=\left(\begin{array}{cc} 0 & {\bar\sigma}^\mu \\ \sigma^\mu &0 \end{array}\right)$
and $\gamma^5=\left(\begin{array}{cc} 1 & 0 \\ 0 & -1  \end{array}\right)$.
Here ${\bar\sigma}^\mu=(1,-\vec\sigma)$ and $\sigma^\mu=(1,\vec\sigma)$.}
\begin{eqnarray}
{\cal L}_{\rm magnetic}&=&  \mu_m^{\rm ano}\left({1\over 2}
 \bar B \sigma^{\mu\nu}\sigma_3 B \right)F^{\rm em}_{\mu\nu}\,,\nonumber\\
{\cal L}_{\rm electric}&=&  {d}_e
\left(-{i\over 2}\bar B \sigma^{\mu\nu}\gamma^5 \sigma_3 B \right)F^{\rm em}_{\mu\nu}\,,
\label{dipoles}
\end{eqnarray}
where $\sigma_3$ acts on the isospin index of the nucleon doublet $B=(p,n)^T$, and
the anomalous magnetic dipole moment $\mu_m^{\rm ano}$ as well as our CP-violating
electric dipole moment ${d}_e$ are given by
\begin{eqnarray}
\mu_m^{\rm ano}&=& e \cdot(-1)\cdot(2D)\int_0^{z_m} dz \left[{1\over z^3}
 f_{1L}^{(0)}f_{2L}^{(0)}\right]\,,\nonumber\\
{d}_e &=& e\cdot (\alpha-\beta)\cdot(2D)\int_0^{z_m} dz
\left[{1\over z^3} f_{1L}^{(0)}f_{2L}^{(0)}\right]\,,
\label{momentvalues}
\end{eqnarray}
where ${1\over z^3}$ factor in the integrals is traced back to $\sqrt{G_5}={1\over z^5}$ and
the curved space gamma matrix carries an extra factor of $z$ since
 the inverse vielbein $e^M_{A}={ z}\delta^M_{A}$.
The coefficient $D$ must be determined by comparing with the experiments
$\mu_m^{\rm ano}=1.8 \mu_N={1.8 e\over 2m_N}$, but for our purpose of obtaining
${d}_e$, we have a model-independent prediction for the ratio
\be
{{d}_e \over \mu_m^{\rm ano}}=-(\alpha-\beta)=-{g \over 4}{m\over m_N}\bar\theta\quad,\label{indratio}
\ee
between the electric dipole moment and the anomalous magnetic dipole moment of nucleons.
This gives us
\be
{d}_e\,=\,-1.8 \mu_N\cdot {g \over 4}{m\over m_N}\bar\theta\, =\,-{1.8g\over 8}\left({m\over m_N}\right)
\left({e\over m_N}\right)\bar\theta\,.
\ee
Using $m_N=0.94$ GeV and ${\rm GeV}^{-1}=0.197\times 10^{-13}\,{\rm cm}$, we have
\be
{d}_e\,=\,-0.25\, g \times \left(m\over 5\,{\rm MeV}\right)
\times 10^{-16}\,\, \bar\theta\quad (e\cdot {\rm cm})\,,
\ee
in units of $(e\cdot {\rm cm})$. This is the main result of our paper.

For a model in~\cite{EKSS} with $z_m=(330\,{\rm MeV})^{-1}$,
$\sigma=(311\,{\rm MeV})^3$ and $m=2.34$ MeV, we have $g=9.18$ to
have the correct nucleon mass eigenvalue $m_N=0.94$ GeV \cite{Ho},
and this predicts that \be {d}_e=-1.08\times 10^{-16}\,\,
\bar\theta\quad (e\cdot {\rm cm})\,. \ee
This means that the neutron electric dipole moment is
${d}_n=-{d}_e=+1.08\times 10^{-16}\,\,
\bar\theta\quad (e\cdot {\rm cm})$.

We end this section by pointing out
a new feature that our result predicts for the first time.
The isospin structure in the operator (\ref{dipoles})
without $U(1)_B$ component tells us that the proton electric dipole moment ${d}_p$
should be equal in magnitude but minus to  the neutron electric dipole moment, that is,
${d}_p=-{d}_n={d}_e$.
It would be very interesting to test this prediction in the future experiments,
which would  confirm the validity of our 5D model of nucleons.

\section{CP-Violating Pion-Nucleon Coupling \label{secv}}

We  calculate the CP-violating component in the coupling of the pions to the nucleons
\be
{\cal L}_{CPodd}=\bar g_{\pi NN}\, (\,\bar B \,\vec\sigma\, B\,)\cdot \vec\pi\quad,
\label{cpodd}
\ee
where $B=(p,n)^T$ is the nucleon iso-doublet and $\vec\sigma$ acts on the
isospin index, while the usual
CP-even interaction is defined  as
\be
{\cal L}_{CPeven}= i g_{\pi NN}\, (\,\bar B \,\vec\sigma\,\gamma^5\, B\,)\cdot \vec\pi\quad.
\ee
As we have the nucleon wave functions $f_{(1,2)(L,R)}$ in the previous section,
what remains is to identify the pion profile along the fifth direction, so that
the above 4D interactions can be read off by integrating out 5D operators over the fifth dimension.
We will pay a special attention to how the strong CP-violation we introduce in
$\left<X\right>=[v_0(z)+i\delta v(z)]{\bf 1}$ induces
a mixing between the CP-even pions and the isoscalar mesons from $X$, since this
will also contribute to (\ref{cpodd}) eventually.

Normally, the pions are identified as the axial $SU(2)_A$ phase angles of the chiral condensate, that is,
$X\sim \left<X\right>e^{i\pi(x)f(z)}$ with a 5D profile $f(z)$.\footnote{
This corresponds only to the phase of the chiral condensate, without affecting
the current quark mass. The reason is that $f(z)$ is normalizable and decays to zero in UV region, and
the IR profile of $X$ encodes the chiral condensate, while its non-normalizable
UV mode contains the current quark
mass. Therefore, the normalizable pions with $f(z)$-profile do not contain the
phase of the current quark mass.}
 In the AdS/QCD set-up, since we
elevate the axial $SU(2)_A$ to a gauge symmetry,
this is no longer a gauge invariant statement, and
we should consider the broken 5D $SU(2)_A$ gauge field simultaneously with the
phase angles of $\left<X\right>$
which could be viewed as eaten Goldstone bosons by the (5D) Higgs mechanism.
In our analysis, we choose to work in the unitary gauge, or equivalently a 5D
version of the $R_\xi$-gauge in $\xi\to \infty$ limit, which will become clear in a moment.
The advantage of this gauge-fixing is an unambiguous diagonalization of
the kinetic terms of the relevant fields, which makes finding equations of motion a lot easier
than other gauge choices.

A modification appearing with the strong CP-violation
turns out to be that not only the phases but also the modulus
of the chiral condensate takes part in the pion wave-functions; in other words,
expanding $X\sim \left<X\right>e^{-Q+iP}$ with $P$ and $Q$ being Hermitian,
there is a small CP-violating mixing between the modulus $Q$ and the $SU(2)_A$ gauge field $A_z$
in addition to the usual mixing of $P$ and $A_z$.\footnote{
There does not appear any CP-violating mixing between $P$ and the vector $SU(2)_V$ since, in our set-up,
$\left<X\right>$ is proportional to identity matrix $\left<X\right>=v(z){\bf 1}=[v_0(z)+i\delta v(z)]{\bf 1}$,
and the $SU(2)_V$ remains unbroken. This may not hold in general for an isospin asymmetric
quark mass $m_u\ne m_d$.}
  ($A_z$ is the component
of the $SU(2)_A$ gauge field along the fifth direction $z$).
This implies that the resulting pion wave-function along the fifth direction
involves a small $Q$-component as well as the dominant $A_z$ and $P$ part.
The CP-violating $Q$-component in the pion wave-function will then contribute
to the CP-violating coupling to the nucleons in (\ref{cpodd}).
Of course, the operator (\ref{cpodd}) also receives contributions from the CP-violating
phase $(\alpha-\beta)$ in the nucleon wave-functions $f_{(1,2)(L,R)}$ that we identify in the previous section.

It is rather straight-forward but tedious to
perform the above mentioned procedure to obtain the 5D profile of the pions.
Defining the axial vector field by $A\equiv {1\over 2}(A_L-A_R)$ and expanding
$X=\left<X\right>e^{-Q+i P}=[v_0+i\delta v]e^{-Q+i P}$ up to the quadratic order,
we have
\begin{eqnarray}
S^{(2)}_{5D}&=&\int d^4x dz \sqrt{G_5} {\rm Tr}
\Bigg\{-{1\over 2g_5^2}(F_A)_{MN}(F_A)^{MN}+(v_0)^2 \Big(2A_M -\partial_M P\Big) \Big(2A^M -\partial^M P\Big)\nonumber\\
&+&(v_0)^2
(\partial_M Q)(  \partial^M Q)
+2\Big((\partial_z v_0) (\partial^z v_0)+3 (v_0)^2 \Big)Q^2
+4v_0(\partial_z v_0) Q (\partial^z Q)\nonumber \\
&-& 4 \Big((\partial_z v_0)\delta v - v_0(\partial_z \delta v)\Big)Q \Big(2A^z-\partial^z P\Big)
\Bigg\}\quad,\label{quadratic}
\end{eqnarray}
where $F_A$ is the field strength of $A$ and the last line is the CP-violating mixing
in the linear order of $\delta v$.
In the above, we should remember that
$\sqrt{G_5}={1\over z^5}$ and $\partial ^z=g^{zz}\partial_z=-z^2 \partial_z$, $A^z=-z^2 A_z$.
As we are interested in the effects from the small CP-violating $\delta v$ in the last line,
we first solve the top line for $A_z$ and $P$ to find the zeroth order profile of the pion,
and then treat the last line as a perturbation to calculate a small induced $Q$-component in the pion wave-function.

The kinetic terms are mixing among $A_\mu$, $A_z$ and $P$ in the first line of (\ref{quadratic}),
and to remove those mixing terms, it is convenient to introduce a 5D version of $R_\xi$
gauge-fixing term~\cite{Ho},
\be
S_{g.f.}=-{1\over 2 \xi }\int d^4x dz \,\,{1\over  z}\Bigg[\partial_\mu A_\mu
-2\xi\Bigg({1\over g_5^2}z\partial_z\left(A_z\over z\right)-{2(v_0)^2\over z^2}P\Bigg)\Bigg]^2\quad,
\ee
where everything is written in the flat metric basis, $ds^2=\eta_{\mu\nu}dx^\mu dx^{\nu}-dz^2$.
We then have separation between $A_\mu$ and $(A_z,P)$, and, since the pions arise from the latter,
we  focus on the $(A_z,P)$ sector only from now on.
We further take $\xi\to \infty$ limit to work in the unitary gauge for simplicity.
This means that we need to impose a constraint for the propagating fields in such a limit,
\be
{1\over g_5^2}z\partial_z\left(A_z\over z\right)={2(v_0)^2\over z^2}P\,,
\label{constraint}
\ee
so that $P$ can be replaced by $A_z$ in the action. We then solve the resulting
action for $A_z$ to find the pion profile.
The resulting relevant action is
\begin{eqnarray}
\int d^4x &dz &{\rm Tr}\Bigg\{
{1\over g_5^2 z}(\partial_\mu A_z)^2 +{z^3\over 4g_5^4(v_0)^2}
\Bigg(\partial_\mu\partial_z\Big({A_z\over z}\Big)\Bigg)^2
-{(v_0)^2\over z^3}\Bigg(2A_z-\partial_z\Big({z^3\over 2g_5^2 (v_0)^2}\partial_z\Big({A_z\over z}\Big)\Big)\Bigg)^2\nonumber\\
&+& {(v_0)^2\over z^3}\Big((\partial_\mu Q)^2-(\partial_z Q)^2\Big)
-{2\over z^3}\Big((\partial_z v_0)^2-{3(v_0)^2\over z^2}\Big) Q^2-{4\over z^3}v_0(\partial_z v_0)  Q\partial_z Q\nonumber\\
&+& {4\over z^3}\Big((\partial_z v_0)\delta v - v_0(\partial_z \delta v)\Big)Q
\Bigg(2A_z-\partial_z\Big({z^3\over 2g_5^2 (v_0)^2}\partial_z\Big({A_z\over z}\Big)\Big)\Bigg)\Bigg\}\quad,
\label{5daction}
\end{eqnarray}
where we first solve the top line and then solve for $Q$ from the last two lines.

The equation of motion can be easily derived from the above. As we are looking for the pion profile,
we put $A_z=\pi(x)A(z)$ and $Q=\pi(x)\,q(z)$
with $q(z)$ being expected to be small  and linear in $\delta v$ or $\bar\theta$\,.
The 4D pion field $\pi(x)$ satisfies
$\partial_\mu\partial_\mu \pi(x)=-m_\pi^2 \pi(x)$ where $m_\pi$
is the pion mass that must be determined from the eigenvalue equation we will specify in a moment.
The equation for $A(z)$ turns out to have a nice factorization structure, due to
 the underlying gauge invariance. To be explicit,
defining
\be
B \equiv 2A -\partial_z\Big({z^3\over 2g_5^2 (v_0)^2}\partial_z\Big({A\over z}\Big)\Big),
\label{btoa}
\ee
we find the equation for $A(z)$ becomes a equation for $B(z)$ only,
\be
\partial_z\Bigg({z^3\over (v_0)^2}\partial_z\Big({(v_0)^2\over z^3} B\Big)\Bigg)
-{4g_5^2 (v_0)^2\over z^2} B = -m_\pi^2 B\quad, \ee with the IR
boundary condition $B(z_m)=0$. The normalizability at UV, $z\to
0$, determines the pion mass $m_\pi$ as the eigenvalue of the
above equation. For example, the parameters in~\cite{EKSS},
$z_m=(330\,{\rm MeV})^{-1}$, $g_5^2=4\pi^2$, and $v_0(z)={1\over
2}(m z+\sigma z^3)$ with $m=2.34$ MeV, $\sigma=(311 \,{\rm
MeV})^3$, indeed give us $m_\pi=140$ MeV as the lowest
eigenvalue~\footnote{One can also easily check that the
Gell-Mann-Oakes-Renner relation holds to a good degree in our
model by calculating the pion mass for different values of the
quark mass.}. The normalization of $B(z)$ should be determined
such that the 4D action of $\pi(x)$ after integrating over the
$z$-direction must have the form \be \int d^4 x \, {\rm Tr} \Big[
(\partial_\mu \pi)(\partial^\mu \pi)- m_\pi^2 \pi^2 \Big]\quad.
\ee Looking at the last term in the first line of
(\ref{5daction}), which produces the mass term \be -\int_0^{z_m}
dz\, {(v_0)^2\over z^3}B^2\cdot \int d^4x \, {\rm Tr} [
\pi^2]\quad, \ee we see that the normalization of $B(z)$ must be
fixed by \be \int_0^{z_m} dz\, {(v_0)^2\over z^3}B^2 \,\,=\,\,
m_\pi^2\quad. \ee Once we solve $B(z)$, we then find $A(z)$ via
solving (\ref{btoa}) with the IR boundary condition $A(z_m)=0$ and
the normalizability in the UV region, $z\to 0$. Finally, $p(z)$ in
$P=p(z)\pi(x)$ is determined through the constraint
(\ref{constraint}), \be {1\over g_5^2}z\partial_z\left(A\over
z\right)={2(v_0)^2\over z^2}p\,. \label{constraint2} \ee

Having obtained the zeroth order profile $A(z)$ (and $p(z)$) for the pion, we next solve $q(z)$
in $Q=q(z)\pi(x)$ that is induced by the CP-violating mixing in the last line of (\ref{5daction}).
Note that the effect of this mixing to $A(z)$ and $p(z)$
 is of second order in $\delta v$, and can be neglected.
The equation for $q(z)$ is written as
\begin{eqnarray}
&&\partial_z\Big({(v_0)^2\over z^3}\partial_z q\Big)
+\Bigg({m_\pi^2 (v_0)^2 \over z^3} -{2\over z^3}\Big((\partial_z v_0)^2-{3(v_0)^2\over z^2}\Big)
+\partial_z\Big({2\over z^3}v_0\,\partial_z v_0\Big)\Bigg)q\nonumber\\
&=& -{2\over z^3}\Big(\delta v\,\partial_z v_0 - v_0\,\partial_z \delta v\Big)B\quad,
\end{eqnarray}
where the second line is the source for $q(z)$ induced by $\delta v(z)={m\bar\theta\over 4}z$.
We should impose the IR boundary condition $q(z_m)=0$ and the UV normalizability as well.

After getting the pion profile $A_z=A(z)\pi(x)$, $P=p(z)\pi(x)$ and $Q=q(z)\pi(x)$,
we see that it is
straightforward to insert them together with the nucleon wave-functions $f_{(1,2)(L,R)}$
 into our 5D action of holographic baryons, and to read off the couplings between pions and the
 nucleons.
Writing the resulting 4D interaction terms as
\be
 {\cal L}_{\pi NN}=i\, g_{\pi NN}^{(1)}\, (\,\bar B \,\vec\sigma\,\gamma^5\, B\,)\cdot \vec\pi
-{g_{\pi NN}^{(2)}\over2 m_N}\, (\,\bar B
\,\vec\sigma\,\gamma^\mu\gamma^5\, B\,)\cdot\partial_\mu\vec\pi
+\bar g_{\pi NN}\, (\,\bar B \,\vec\sigma\, B\,)\cdot
\vec\pi\quad, \ee where we expand
$\pi=\vec\pi\cdot{\vec\sigma\over 2}$, the last term is the
CP-violating pion-nucleon coupling we are interested in, with the
coupling strength $\bar g_{\pi NN}$ given by
\begin{eqnarray}
\bar g_{\pi NN} &=& -(\alpha-\beta)\int_0^{z_m}dz\, \Bigg[{1\over z^4}A(z)f_{1L}^{(0)}(z)f_{2L}^{(0)}(z)
+{g\over 2z^5} v_0(z)p(z)\Big[(f_{1L}^{(0)}(z))^2+(f_{2L}^{(0)}(z))^2\Big]\Bigg]\nonumber\\
&+& {g\over 2} \int_0^{z_m} dz\, \Bigg[{1\over z^5}\Big(v_0(z)q(z)+\delta v(z)p(z)\Big)
\Big[(f_{1L}^{(0)}(z))^2-(f_{2L}^{(0)}(z))^2\Big]\Bigg]\quad,
\end{eqnarray}
where
$(\alpha-\beta)={g \over 4}{m\over m_N}\bar\theta$ and $\delta v(z)={m\bar\theta\over 4}z$
as given before. This is the primary result of this section.
Since $q(z)$ is proportional to $\bar\theta$ via $\delta v$,
 our $\bar g_{\pi NN}$ is proportional to the strong CP angle $\bar\theta$ as expected.

For the model of $z_m=(330\,{\rm MeV})^{-1}$ and $v_0(z)={1\over
2}(m z+\sigma z^3)$ with $m=2.34$ MeV, $\sigma=(311 \,{\rm
MeV})^3$, we have $g=9.18$ and our numerical result for $\bar
g_{\pi NN}$ is \be \bar g_{\pi NN}\,\,=\,\,+0.017
\,\,\bar\theta\quad, \ee which is about half from the previous
estimate in~\cite{Crewther:1979pi}.

For completeness, we also give the expressions for the CP-conserving
coupling constants $g_{\pi NN}^{(1)}$ and $g_{\pi NN}^{(2)}$,
\begin{eqnarray}
g_{\pi NN}^{(1)} &=& \int_0^{z_m} dz\, \Bigg[{1\over z^4}A(z)f_{1L}^{(0)}(z) f_{2L}^{(0)}(z)
+{g\over 2 z^5} v_0(z)p(z)\Big[(f_{1L}^{(0)}(z))^2+(f_{2L}^{(0)}(z))^2\Big]\Bigg]\quad,\nonumber\\
g_{\pi NN}^{(2)} &=& -(2m_N)\cdot D \int_0^{z_m} dz\, \Bigg[{1\over
z^3} A(z)\Big[(f_{1L}^{(0)}(z))^2+(f_{2L}^{(0)}(z))^2\Big]\Bigg]
\quad.
\end{eqnarray}
Note that the coupling term with $g_{\pi NN}^{(2)}$ may turn into
the $g_{\pi NN}^{(1)}$-coupling by using the equation of motion
for $B$, and the total pion-nucleon coupling is $g_{\pi NN}=g_{\pi
NN}^{(1)}+g_{\pi NN}^{(2)}$. Recall also that $D$ should be fitted
against the anomalous magnetic dipole moments of the proton and
neutron by (\ref{momentvalues}), \be (2D)\cdot\int_0^{z_m} dz
\left[{1\over z^3}
f_{1L}^{(0)}f_{2L}^{(0)}\right]\,\,=\,\,-{\mu_m^{\rm ano}\over
e}\,\,\simeq\,\,-{1.8 \over 2 m_N}\quad. \ee Numerical values for
the above model are $g_{\pi NN}^{(1)}=-3.64$, $g_{\pi
NN}^{(2)}=-22.0$ with $g_{\pi NN}=-25.6$, which is about twice of
the experiment $g_{\pi NN}^{\rm exp}=13.6$. There is however much
chance to improve several crude simplifications in constructing
our model, such as the simple AdS-slicing, neglecting anomalous
dimension of the three-quark baryon operators, the possibility of
a different IR-cutoff, {\it etc}.~\cite{Evans:2005ip}.

\section{Discussion and Conclusion}

We have studied the CP-violating effects in strong interactions, induced by the
QCD $\theta$-term,
in a bottom-up model of holographic baryons~\cite{Ho}. We find that the model of holographic
baryons is quite useful in studying the CP-violating effects in the baryon sector such as
the electric dipole moments of nucleon or the CP-violating couplings of pions
to nucleons, since the CP-violating effects are
easily incorporated in the bulk scalar field $X$, whose coupling to bulk spinors
naturally leads to the calculable CP-violating effects in the baryon sector.

A Pauli term is added to the  holographic model of
baryons~\cite{Ho} to estimate the electric dipole moments of
nucleons. We have fixed the coefficient of the Pauli term by
fitting the anomalous magnetic moments of nucleons, though in
top-down holographic models such as Sakai-Sugimoto
model~\cite{Sakai:2004cn} the coefficient can be calculated
reliably due to the fact that the baryons are realized as
instanton solitons in holographic QCD~\cite{HRYY,Hata:2007mb}.
However, we stress that our result for the electric dipole moment
of nucleons is model-independent, since it relies only on the
ratio between the electric dipole moment and the anomalous
magnetic moment of nucleons, $d_e/ \mu_m^{\rm ano}\,$, which is
independent of the coefficient of the 5D Pauli term, as shown in
(\ref{indratio}).

In fact, we can argue further that our result is universal against any
other possible higher dimensional
operators in 5D. Note that in solving (\ref{eqn}), we can alternatively go to the frame where the
nucleon wave-functions $f_{(1,2)(L,R)}$ are real, whereas the strong CP-violation
is encoded instead in the phase of the nucleon mass $m_N\to e^{i(\alpha-\beta)}m_N$.
Upon reduction to 4D, all the resulting 4D operators obtained from real wave functions
$f_{(1,2)(L,R)}$,
including anomalous magnetic dipole moment operator,
will be CP-conserving except the nucleon mass term with the complex axial mass.
Performing an axial $U(1)_A$ rotation back to make the nucleon mass real,
this will induce the electric dipole moment operator from the anomalous magnetic moment operator
as in (\ref{dipoles}) with the universal prediction for the ratio
\be
{{d}_e \over \mu_m^{\rm ano}}=-(\alpha-\beta)=-{g \over 4}{m\over m_N}\bar\theta\,,
\ee
without any regard to the details of the 5D model.

Though our model is bottom-up, the bulk spinors in our model
should be interpreted as effective fields of the instanton
solitons as in the top-down model, because the mesonic sector of
our model, being a 5D gauge theory with a Chern-Simons term,
naturally supports instanton solitons, whose Wilson line can be
identified as
skyrmions~\cite{Atiyah:1989dq,Son:2003et,Nawa:2006gv}. As shown
in~\cite{HRYY}, the Pauli term therefore should not contain the
$U(1)$ gauge fields to reproduce the correct behavior of
non-abelian instanton solitons at long distances, which leads to
important sum rules for both the electric dipole moments and the
anomalous magnetic moments of nucleons, which are
model-independent and not subject to any $1/N_c$ corrections,
\begin{equation}
d_p+d_n=0,\quad \mu^{\rm ano}_p+\mu^{\rm ano}_n=0\,.
\end{equation}

Combining our holographic estimate of the neutron electric dipole
moment, given as $d_n=1.08\times 10^{-16}\,\, \bar\theta\, e\cdot
{\rm cm}\,$, with the current experimental bound on the neutron
EDM~\cite{Baker:2006ts}, we obtain a limit on the $\theta$ angle
reliable up to $30\%$ coming from $1/N_c$ corrections,
\begin{equation}
\left|\bar\theta\right|\,<3\times10^{-10}\,,
\end{equation}
which is somewhat stronger than the previous bounds.

%\vskip 1in
%1. generality of our result : going to the complex mass + real eigenfunctions basis

%2. adding contact term + chiral log enhancement

\vfill \eject

\acknowledgments
We are grateful to Jihn E. Kim and Elias Kiritsis for useful discussions, and thank Stanley Brodsky
for letting us know Ref.\cite{Brodsky:2006ez}.
One of us (H.U.Y.) also thanks Kimyeong Lee for a financial support from his research fund.
This work was supported in part (D.K.H.) by KOSEF
Basic Research Program with the grant No. R01-2006-000-10912-0,
(H.U.Y.) the Korea Research Foundation Grant KRF-2005-070-c00030, and
(H.Ch.K.) Korea Research Foundation Grant funded by the Korean
Government(MOEHRD) (KRF-2006-312-C00507).
The work of S.~S. was supported by the 2nd stage of Brain Korea 21 project for the year 2007.

\end{document}